\documentclass [superscriptaddress,amsmath,prl,twocolumn,amssymb]{revtex4-1}
\usepackage{graphicx}% Include figure files
\usepackage{dcolumn}% Align table columns on decimal point
\usepackage{bm, color}
\usepackage[none]{hyphenat} % avoid the hyphen "-" that splits a word in the end of a line.
\usepackage{bbm}
\usepackage{braket}
\usepackage{color}% label text with different colors

\begin{document}
\title{Prediction of triple point fermions in simple half-Heusler topological insulators}

\author{Hao Yang}
\affiliation{Max Planck Institute of Microstructure Physics, Weinberg 2, 06120 Halle, Germany}
\author{Jiabin Yu}
\affiliation{Department of Physics, the Pennsylvania State University, University Park, PA, 16802}
\author{Stuart S. P. Parkin}
\affiliation{Max Planck Institute of Microstructure Physics, Weinberg 2, 06120 Halle, Germany}
\author{Claudia Felser}
\affiliation{Max Planck Institute for Chemical Physics of Solids, 01187 Dresden, Germany}
\author{Chao-Xing Liu}
\affiliation{Department of Physics, the Pennsylvania State University, University Park, PA, 16802}
\author{Binghai Yan}
\email{binghai.yan@weizmann.ac.il}
\affiliation{Department of Condensed Matter Physics, Weizmann Institute of Science, Rehovot, 7610001, Israel}

\begin{abstract}
We predict the existence of triple point fermions in the band structure of several half-Heusler topological insulators by $ab~initio$ calculations and the Kane model.
We find that many half-Heusler compounds exhibit multiple triple points along four independent $C_3$ axes,
through which the doubly degenerate conduction bands and the nondegenerate valence band cross each other linearly nearby the Fermi energy.
When projected from the bulk to the (111) surface, most of these triple points are located far away from the surface $\bar{\Gamma}$ point, as distinct from previously reported triple point fermion candidates. These isolated triple points give rise to Fermi arcs on the surface, that can be readily detected by photoemission spectroscopy or scanning tunneling spectroscopy.
\end{abstract}

\maketitle
The discovery of topological insulators (TIs) ~\cite{Qi2011RMP,Hasan2010RMP} has generated much interest in the search for other novel topological states in condensed matter physics and materials science. As quasiparticle analogues of elementary particles of the standard model, Dirac fermions~\cite{Young2012,Wang2012,Liu2014Na3Bi,Xu2015} and Weyl fermions~\cite{Wan2011,volovik2003universe,Burkov2011,Weng2015,Huang2015,Lv2015TaAs,Xu2015TaAs,Yang2015TaAs} have recently been found in several materials (see reviews Refs.~\onlinecite{Yan2017,Armitage2017}).  
Both Dirac and Weyl fermions exhibit Fermi arcs, unclosed Fermi surfaces, on the boundary, as a hallmark for the experimental detection. More recently, several exotic types of fermions, which do not have elementary particle counterparts, have been theoretically predicted as quasiparticle excitations near certain band crossing points that are protected by specific space-group symmetries~\cite{Wieder2016,Bradlyn2016}. In particular, triple point (TP) fermions have been predicted in many materials with triply degenerate band crossing points~\cite{Zaheer2013,Zhu2016,Winkler2016,Weng2016TaN,Weng2016,Yu2017,Fulga2017}.  These predictions have stimulated intensive experimental studies to search for their signatures, for example, using angle-resolved photoemission spectroscopy (ARPES)~\cite{Lv2016} and transport properties~\cite{Shekhar2017}.

TPs can be viewed as an intermediate phase between fourfold degenerate Dirac points and twofold degenerate Weyl points. They also give rise to Fermi arcs when projected onto certain specific crystal facets. However, the detection of TP-induced Fermi arcs remains challenging from the material point of view. A pair of TPs are protected by \textcolor{black}{ the $C_{3v}$ symmetry group (generated by a $C_3$ rotation and a $\sigma_v$ mirror operation)} in certain compounds~\cite{Zaheer2013,Zhu2016,Winkler2016,Weng2016TaN,Weng2016,Yu2017}, for example, tensile-strained HgTe~\cite{Zaheer2013},
 MoP~\cite{Zhu2016}, and antiferromagnetic (AFM) half-Heusler compounds  (e.g. GdPtBi)~\cite{Yu2017}. Even presuming that samples can be grown, the natural cleavable surface is usually the facet that is perpendicular to the $C_3$ axis. Consequently, two TPs at the unique $C_3$ axis are projected to the same $\bar{\Gamma}$ point of the surface Brillouin zone (BZ), resulting in the disappearance of Fermi arcs, as shown in a recent ARPES measurement on MoP~\cite{Lv2016}. Therefore, TP materials with easily measurable Fermi arcs are still required for the final experimental verification of TP fermions.

In this work, we predict the existence of multiple TPs in several half-Heusler compounds in which the detection of Fermi arcs by ARPES and other surface sensitive techniques such as scanning tunneling spectroscopy(STS) should be straightforward.
The face-centered-cubic lattice of half-Heusler compounds has four equivalent $C_3$ axes (e.g. the [111] axis) and, thus, can host four (or multiples of four) pairs of TPs. When projected onto the (111) surface, an easily cleavable plane ~\cite{liu2011,Liu2016Heusler}, TPs at the [111] axes merge into the surface $\bar{\Gamma}$ point while the other three (or a multiple of three) pairs of TPs appear away from $\bar{\Gamma}$, leading to Fermi arcs that link these individual TPs on the surface. Combining $ab~initio$ band structure calculations and the $k \cdot p$ Kane model, we predict several TP candidate half-Heusler materials, including, for example, YPtBi, LuPtBi, and GdPtBi [the paramagnetic phase]). The TPs and resultant extended Fermi arcs are revealed in our calculations, and await experimental proof.

Ternary half-Heusler compounds have been extensively studied in the search for TIs~\cite{chadov2010,Lin2010,Xiao2010Heusler,Yan2014,liu2011,Liu2016Heusler,Logan2016} and Weyl semimetals~\cite{Hirschberger2016,Shekhar2016,Ruan2016,Cano2017}.
The band structure of Heusler TIs has been identified as being topologically identical to HgTe~\cite{Bernevig2006d}. For example, the conduction and valence bands touch each other at the $\Gamma$ point, where the wavefunctions are comprised mainly of $p$-orbitals and are, therefore, named $\Gamma_8$ bands with a total angle momentum $J=3/2$, as shown in Fig. 1a. The fourfold degeneracy at the $\Gamma$ point is protected by time-reversal symmetry (TRS) and $T_d$ group symmetry. 
The $s$-type $\Gamma_6$ bands ($J=1/2$) located below $\Gamma_8$, thus give rise to an inverted band structure.
Along each $C_3$ axis (e.g. the [111] direction),  $\Gamma_8$ bands split into one doubly degenerate band (labelled as $\Lambda_6$ according to the $C_{3v}$ symmetry) and two non-degenerate bands (labelled as $\Lambda_{4, 5}$) due to the absence of inversion symmetry in $T_d$ group. The $\Lambda_6$ bands cross the $\Lambda_{4, 5}$ bands since $\Lambda_6$ and $\Lambda_{4, 5}$ bands disperse oppositely for large $k$ (see Fig. 1a). As already pointed out in  Ref.~\cite{Zaheer2013}, TPs exist at the crossing point between $\Lambda_6$ and $\Lambda_4$ (or $\Lambda_5$) bands.
Unfortunately, these TPs that are located extremely close to the $\Gamma$ point ($\sim$ 0.8\% of the $\Gamma-L$ distance)~\cite{Zaheer2013} and cannot be resolved by currently available techniques.
In contrast, TPs in some Heusler materials can be pushed to very large momenta, because their $\Lambda_6$ bands exhibit a peculiar double-valley feature that is absent in HgTe (e.g. see Fig. 1b).
As illustrated in Fig. 1\textcolor{black}{b}, a pair of TPs may exist near the $\Gamma$ point along the $C_3$ axis ($\Gamma-L$),
where two TPs are \textcolor{black}{related} by TRS and protected by $C_3$ rotational symmetry. Given four $C_3$ axes, four pairs of TPs form inside the first bulk BZ. When projected onto the (111) surface, three pairs are isolated from each other and are far from the surface $\bar{\Gamma}$ point, giving rise to Fermi arcs connecting these six TPs (Fig. 1h).

\begin{figure}[t]
\includegraphics[width=\columnwidth]{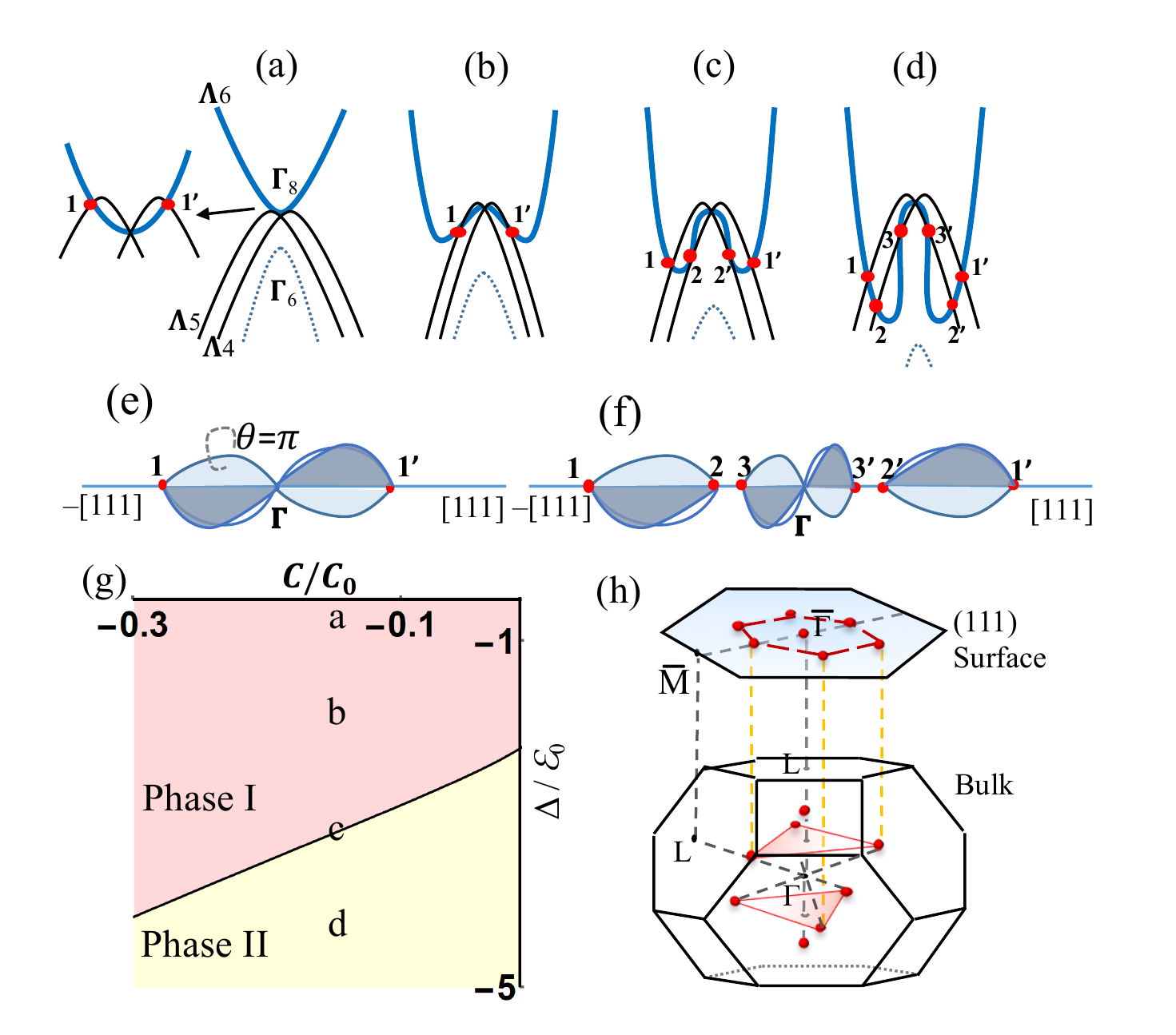}
\caption{
\label{phase_digram}
Evolution of band structures with increasing numbers of triple points (TPs).
(a) HgTe-type band structure along the line $L-\Gamma-L$. The $\Gamma_8$ bands (solids curves) lie above $\Gamma_6$ (dotted curves).  The $\Gamma_8$ bands split into doubly degenerate $\Lambda_6$ (thick solid blue curve) and nondegenerate $\Lambda_4$ and $\Lambda_5$ bands (thin solid black curves) along the $C_3$ axis ($\Gamma-L$). Here $\Lambda_6$ crosses $\Lambda_{4,5}$, forming TPs (filled red circles) very close to the $\Gamma$ point, where \#1 and \#1$^\prime$ represent a TP and its time-reversal partner, respectively.
(b) Heusler-type band structure. The $\Lambda_6$ bands exhibit a double valley shape, pushing a pair of TPs out from the $\Gamma$ point.
(c) Heusler-type band structure with two pairs of TPs along one $C_3$ axis.
(d) Heusler-type band structure with three pairs of TPs along one $C_3$ axis.
(e) Corresponding to the band structures in (a) and (b), three nodal lines (blue curves) inside the $C_{3v}$ mirror planes connect a pair of TPs (1 and $1^\prime$) along the $C_3$ axis by passing the $\Gamma$ point.
(f) Corresponding to the band structure in (d), the nodal lines connects these three pairs of TPs.
(g) Phase digram of TPs with respect to the band inversion strength $(\Gamma_6 - \Gamma8)/\epsilon_0$ and the linear splitting term $C/C_0$ of $\Lambda_{4,5}.$ Within the phase digram, a-d correspond to the band structures of (a)-(d), respectively.
(h) Distribution of TPs in the bulk Brillouin zone and their projection onto the (111) surface.
The surface Fermi arcs that connect different TPs are illustrated by red dashed lines.
}
\end{figure}

We first construct a phase digram to reveal the emergence and properties of TPs in half-Heusler compounds 
based on the Kane model\cite{winkler2003spin}, in order to guide the material search.
The crystal symmetry of Half-Heusler materials is described by the space group $F\bar{4}3m$ and 
the point group $T_d$ ~\cite{canfield1991magnetism}, respectively,
and the corresponding low energy physics can be described by the six-band Kane model with 
two $\Gamma_6$ bands and four $\Gamma_8$ bands. 
Along any of the four $C_3$ axes, two $\Gamma_6$ bands are still degenerate, labeled as $\Lambda_6^-$ bands, 
according to the irreducible representations of $C_{3v}$ spin double group,
while four $\Gamma_8$ bands are split into one doubly degenerate band, denoted as $\Lambda_6^+$ bands, 
and two nondegenerate bands, denoted as $\Lambda_4$ and $\Lambda_5$ bands, respectively. 
%The $\Lambda_6^+$ carries total angular momentum $\pm \frac{1}{2}$ along the $C_3$ direction, known as
%light hole in semiconductor physics, 
%while the $\Lambda_{4,5}$ bands carries $\pm \frac{3}{2}$ total angular momentum, known as
%heavy hole. 
For the momentum close to $\Gamma$ along the $C_3$ axis, 
the $\Lambda_6^+$ bands disperse quadratically while the $\Lambda_{4,5}$ bands disperse linearly
with opposite velocities for two branches due to the linear C term in the Kane Hamiltonian. Thus, 
the $\Lambda_6^+$ bands must locate between two $\Lambda_{4,5}$ bands for small momenta. 
For larger momenta, the $\Lambda_6^+$ bands bend up and thus will be always above two $\Lambda_{4,5}$
bands that bend down. Thus, we conclude
that at least one pair of TPs due to the crossing between the $\Lambda_6$ bands and the upper branch of
$\Lambda_{4,5}$ bands must exist. 
According to energy disperion along the $C_3$ axis for the Kane model (see details in Ref.~\cite{supplementary}), 
we find that 1, 2 or 3 pairs of TPs can exist in one $C_3$ axis, depending on model parameters. 
The phase diagram as a function of the parameter $C$, which determines the energy splitting of 
two $\Lambda_{4,5}$ bands, and the gap $\Delta$ between $\Gamma_6$ and $\Gamma_8$ states, which influence
the effective mass of $\Lambda_6^+$ bands, is shown in Fig. 1g. 
For a small $\Delta$, strong hybridization beween $\Lambda_6^+$ and $\Lambda_6^-$ bands
can lead to a positve effective mass for the $\Lambda_6^+$ band and thus results in 1 pair
of TPs between the $\Lambda_6^+$ bands and the upper branch of $\Lambda_{4,5}$ bands (Fig. 1a)
in the phase I in Fig. 1g. 
As $\Delta$ increases, the effective mass for the $\Lambda_6^+$ bands become
negative, leading to a double-hump structure (Fig. 1b). With $\Delta$ increasings to a critical value, 
the double-hump $\Lambda_6^+$ bands can touch the lower branch of $\Lambda_{4,5}$ bands, giving
rise to one more pair of TPs (TP \#2 and \#2') at the critical line in Fig. 1g. 
As $\Delta$ further increases, the $\Lambda_6^+$ bands
can cross the lower branch of $\Lambda_{4,5}$ twice, resulting in 3 pairs of TPs in total for the phase I in Fig. 1g. 
TPs are connected by nodal lines and for different phases, we find the connections are different. 
For the phase I, four nodal lines, three in three mirror planes and one along the $C_3$ axis,
connects the TP \#1 to its time-reversal partner TP \#$1^\prime$,
passing through the $\Gamma$ point, as shown in Fig. 1e.
The Berry phase around each of three nodal lines in the mirror plane is accumulated to $\pi$
and characterizes its topological nature (Type-B TPs 
\textcolor{black}{introduced} in Ref.~\cite{Zhu2016}). 
Three pairs of TPs exist the phase II, with 
TPs \#1 and \#2 (\#$1^\prime$ and \#$2^\prime$) connected by nodal lines
and TPs \#3 and \#$3^\prime$ connected by nodal lines through $\Gamma$, as shown in Fig. 1f.

\begin{figure}[t]
\includegraphics[width=0.8\columnwidth]{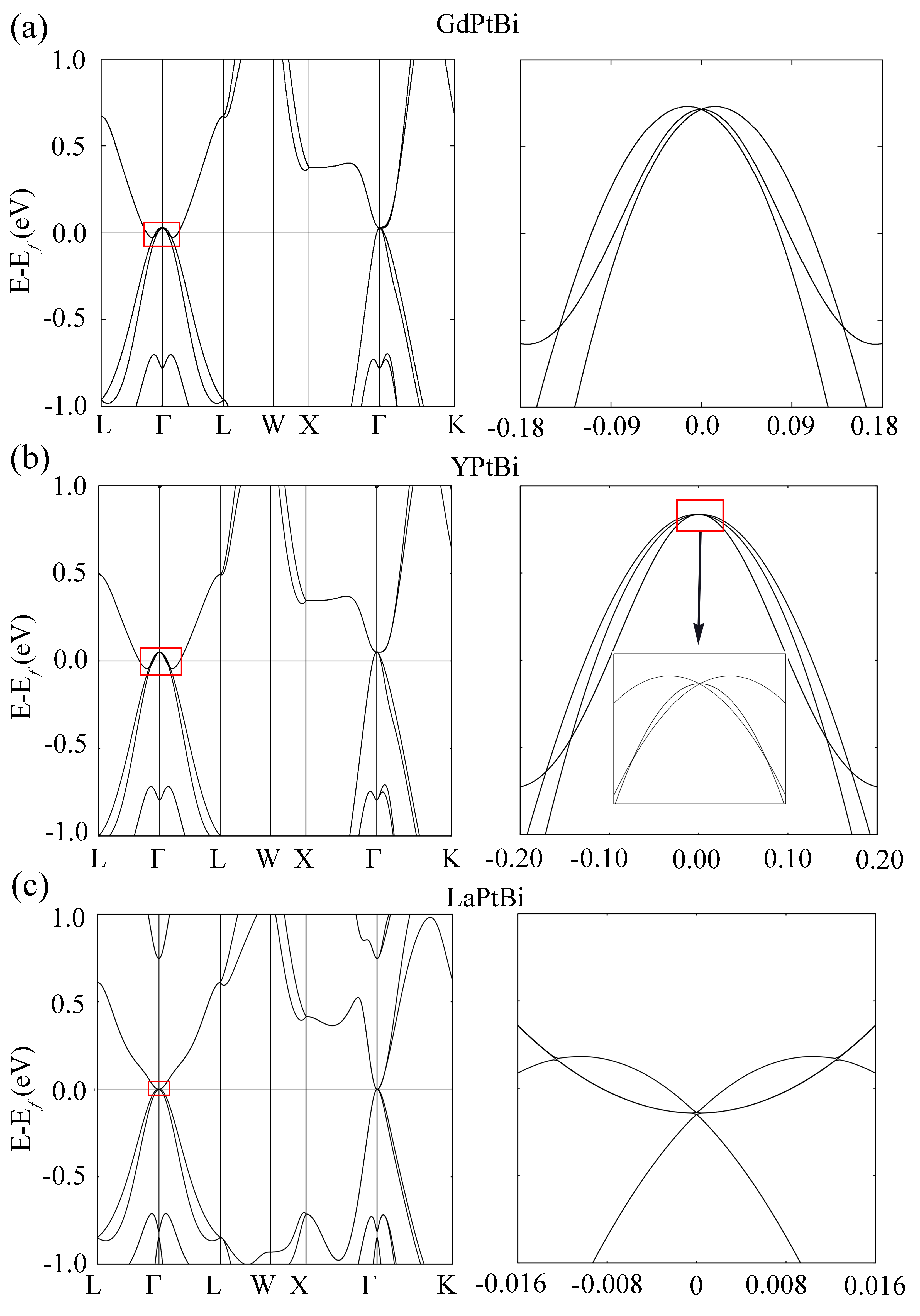}
\caption{
\label{phase_digram}
Bulk band structures with triply degenerate band crossings.
The long-TP materials (a) GdPtBi with 2 TPs along the $C_3$ axis and (b) YPtBi with 6 TPs.
The band dispersion along  the $C_3$ axis near the $\Gamma$ point are magnified to show the TPs on the right panels.
(c) The short-TP material LaPtBi are shown for comparison.
}
\end{figure}

To search for ideal material candidates, we have performed $ab~initio$ band structure calculations for a large number of half-Heusler compounds using density-functional theory (DFT) with the generalized gradient approximation.  We have identified many candidate materials exhibiting TPs with a large momentum separation in their band structure, as listed in Table I. For example, TPs \#1 and \#2 of $R$PtBi ($R=$ Y, Lu), LuAuPb, LuPdBi, and TP \#1 of GdPtBi (paramagnetic phase) lie at large momenta greater than \%10 of the $\Gamma-L$ length.
The TP \#3 located too close to the the $\Gamma$ point for the observation.
We also list the band inversion strength between the $\Gamma_6$ and $\Gamma_8$ bands in table I to demonstrate the  evolution of the band structure.
Roughly consistent with the above phase digram, TPs shift to larger momenta as the band inversion is enhanced.
For comparison, we also show the band structure of LaPtBi in Fig. 2, where TPs appear very close to the $\Gamma$ point.
For convenience, we term materials with TPs at large momenta as long-TP materials, and those with TPs at tiny momenta as short-TP materials.
When the Fermi energy crosses a TP in GdPtBi and YPtBi, the TP behaves as the touching point between hole and electron pockets, thereby showing the same feature as a type-II Weyl semimetal~\cite{Soluyanov2015WTe2,Sun2015MoTe2}.
Many half-Heusler compounds are known to exhibit much larger band inversions than HgTe. Thus, it is not surprising to find long-TP materials here.  It has been reported that optimized exchange-correlation functionals in DFT tends to reduce the band inversion of Heusler compounds~\cite{Feng2010}.
We note that this functional correction remains of the general order of the band inversion strength between different Heuslser materials,  where long-TP materials can still be found in the large band inversion region.

\begin{figure*}[t]
\includegraphics[width=2\columnwidth]{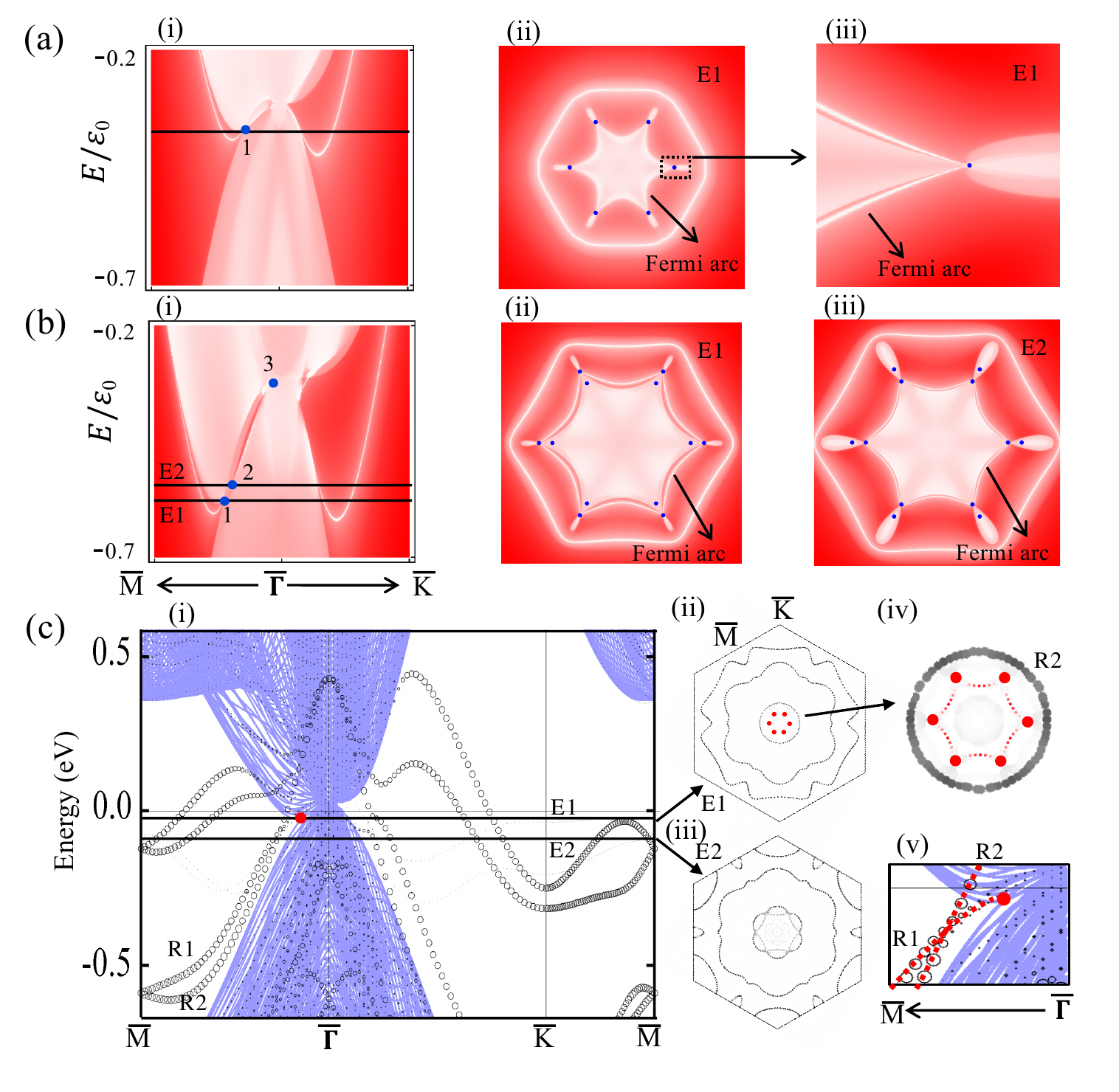}
\caption{
\label{phase_digram}
Surface band structures. (a) and (b), Band structure calculated by the tight-binding model on a half-infinite surface with 1 and 3 TPs  along $\bar{\Gamma}-\bar{M}$, respectively.
Red color stands for low surface intensity and white for strong surface intensity. The triple points are marked by blue points.
(c) DFT band structure calculated on a slab model for GdPtBi (paramagnetic phase).
(i) The size of the white circles represent the surface contribution and, thus, large circles show the surface states.
Bulk bands are indicated by blue curves as a background, where the triple point is shown as the red point.
R1 and R2 are a pair of Rashba-like surface states.
(ii) and (iii), Fermi surfaces corresponding to energy $E1$ (crossing the triple point) and $E2$, respectively.
The flower-like Fermi surface in (iii) was measured in previous ARPES experiments.
(iv) Magnified inner Fermi ring of (ii).
Fermi arcs (red dotted lines) are artificially added as guides to the eye,
for they are missing in the DFT band structure due to the finite slab thickness.
(v) Expanded view near the triple point. Red dotted lines highlight R1 and R2 bands.
The R1 band crosses R2 and later ends at the triple point.
}
\end{figure*}

\begin{table}[htp]
\caption{List of triple point (TP) half Heusler materials. The band inversion strength $\Delta$ is in units of eV.
The number of TPs (TP\#) is shown in Fig. 1.
The distance of a TP to the $\Gamma$ point is specified as a percentage of the $\Gamma-L$ length.
The energy of a TP is given with respect to the Fermi energy.}
\begin{center}
\begin{tabular}{|r|r|r|r|r|}
\hline
Material & TP\#  & $\Delta$ & Distance to $\Gamma$ & Energy (meV)\\
\hline
LuPtBi   & 1 &-1.52 & 32.5\%  & -131 \\
              & 2 &         & 26.7\%   & -144 \\
              & 3 &         &1.3\%     & 154 \\
LuAuPb & 1 & -1.07& 24.2\%  & 105 \\
              & 2 &         &15.9\%  & 142 \\
              & 3 &         & 5.0\%   & 201 \\
YPtBi     & 1 &-1.07 & 17.1\%  & -38 \\
              & 2 &         & 14.3\%  & -27 \\
              & 3 &         & 0.4\%    & 50 \\
GdPtBi   & 1 &-1.02 & 14.1\%  & -22 \\
LuPdBi  & 1 & -0.69 & 11.9\% & -8 \\
LaPtBi   & 1 & -0.82 & 1.3\%   &8 \\
\hline
\end{tabular}
\end{center}
\label{default}
\end{table}%

The existence of Fermi arcs \textcolor{black}{on the surface} is a hallmark of TPs for their experimental detection. When projected to the (111) surface, six TPs  (TPs \#1 and \#$1^\prime$) locate at the $\bar{\Gamma}-\bar{M}$ line. Because a typical TP is equivalent to two degenerate Weyl points with opposite charities,  typically two Fermi arcs are expected to emerge from a TP. 
A natural choice is that these two Fermi arcs end at two neighboring TPs separately (One possible case is that a Fermi arc connects those two WPs and disappear as two WPs merge to be a TP). 
As a consequence, six Fermi arcs form a hexagon-like Fermi surface.
We first employed a tight-binding regularization of the Kane model and calculated the surface states on a half-infinite (111) surface. As shown in the surface band structure of Fig. 3a with only 1 TP along  the $\bar{M}-\bar{\Gamma}$ line,
a surface band disperses from the Brillouin zone boundary to the center.
Along $\bar{K}-\bar{\Gamma}$, it runs very close to the $\bar{\Gamma}$ point and merges into the bulk background.
From $\bar{M}$ to $\bar{\Gamma}$, however, it ends exactly at the TP.
On the Fermi surface at $E_F$ crossing the TP, one can clearly see that six Fermi arcs connect six TPs, forming a hexagon shape.
Each Fermi arc starts at a TP and ends at the neighboring TP (Fig. 3a-iii).  Outside the hexagon of Fermi arcs, there is a larger Fermi ring due to the same surface band.
When 3 TPs exist along the the $\bar{M}-\bar{\Gamma}$ line (Fig. 3b), the original surface band still ends at the TP \#1 while a new surface band appears to link TP \#2 and TP \#3 although it is weak in intensity.
On the Fermi surface, one can observe that six Fermi arcs connect six TPs \#1 (Fig. 3b-ii) and also  six TPs \#2 (Fig. 3b-iii).
We note that the Fermi arc states penetrate deeply into the bulk,
similar to the Fermi arcs of a Weyl semimetal TaAs~\cite{Batabyal2016,Inoue2016}, since they appear close to the bulk pocket boundary on the Fermi surface.
Therefore, we can summarize two important features of TP surface states. (i) The surface band ends at the TP \#1 position in the energy dispersion along $\bar{M}-\bar{\Gamma}$. (ii) Six Fermi arcs interconnect six TPs \textcolor{black}{related by $C_3$ and TRS} when $E_F$ crosses the TPs.

Regarding materials we have calculated the surface states of GdPtBi based on $ab~initio$ DFT calculations within a slab model.
The slab model includes 54 atomic layers of the (111) surface and the band structure is projected to the top surface that is terminated by Bi atoms.
The projected band structure represents the dispersions of surface states (Fig. 3c), which agrees well with previous ARPES measurement~\cite{liu2011}.
There are several trivial surface states due to Bi dangling bonds in the band structure.
We point out a pair of Rashba-like surface bands (noted R1 and R2 in Fig. 3c), which disperse up from -0.6 eV at $\bar{M}$ to above $E_F$ at $\bar{\Gamma}$. When approaching the TP, R1 does not disperse up together with R2, as ordinary Rashba bands do.
Instead, R1 crosses R2 at energy $E2=-72$ meV and then ends at the TP at $E1=-22$ meV (Fig. 3c-v), fulfilling the first feature of Fermi arc states.
Here the surface band structure is a result of the strong hybridization between Fermi arc states and dangling bond states.
We point out that the same feature that R1 ends at the TP can be found for another long-TP material YPtBi and even a short-TP material LaPtBi,
see Ref. ~\cite{supplementary} for more information.
For the Fermi surface corresponding to $E_2$ (Fig. 3c-iii), there are two rings forming a flower like shape caused by the R1-R2 crossing.
For the Fermi surface corresponding to $E_1$ (Fig. 3c-ii), there is only one apparent ring due to R2.
Here Six TPs locate inside the R2 ring, where Fermi arcs are expected to exist.
However, these Fermi arcs are missing in the DFT band structure.
This is due to the finite size effect of the slab model simulations.
In experiment, corresponding Fermi arcs should appear but possibly with weak intensity, since they penetrate deeply into the bulk.
In previous ARPES experiments, a flower-like Fermi surface similar to Fig. 3c-iii was observed for LuPtBi, YPtBi and GdPtBi~\cite{liu2011,Liu2016Heusler}, where R1 starts crossing R2.
However, ARPES did not reach the energy window of TPs, because these Heusler samples are usually hole doped.
To fully reach the TP region by ARPES, electron-doped samples are needed to shift $E_F$ by $\sim 50 $ meV with respect to current samples.

The predicted Heusler TP materials are known to exhibit antiferromagnetism (AFM) (e.g. GdPtBi)~\cite{canfield1991magnetism} and superconductivity (e.g. LuPtBi and YPtBi)~\cite{Butch2011,Tafti2013} at low temperatures. They serve as a new platform to investigate the interplay between TPs and magnetism / superconductivity. For example, GdPtBi is AFM below 9 K and can be polarized to be ferromagnetic (FM) by a strong magnetic field. Several topological phases are anticipated to exist with an AFM or FM, such as the AFM TI~\cite{Mong2010}, Weyl semimetal~\cite{Hirschberger2016,Shekhar2016,Cano2017,Yu2017}, and mirror TI~\cite{Yu2017}.
It is interesting to ask how the TPs in the bulk and corresponding connected Fermi arcs on the surface in paramagnetic GdPtBi
evolve into Weyl points and new Fermi arcs, or Dirac-cone-like surface states in the AFM or FM phase.
In addition, we note that TPs can also exist in the normal zincblende type band structures without a band inversion, since $\Lambda_6$ bands commonly have different mass from $\Lambda_{4,5}$.

In summary, we have predicted the existence of TP fermions in the band structures of several half-Heusler TIs.
By $ab~initio$ calculations and the $k \cdot p$ Kane model, we have identified the existence of multiple TPs at large momenta in the bulk and revealed the existence of Fermi arcs on the surface.
The Fermi arcs states ends at the TP position in the energy dispersion along $\bar{M}-\bar{\Gamma}$ and connect neighboring TPs at the Fermi surface.
To observe TPs and Fermi arcs, currently available samples may need slightly more electron doping for ARPES studies.
Alternatively, two-photon photoelectron spectroscopy or scanning tunneling spectroscopy, which can measure empty states, will be ideal for the detection of TP Fermi arcs.

\begin{acknowledgments}
We thank Haim Beidenkopf, Nurit Avraham, and Ady Stern for helpful discussions.
B.Y. acknowledges support of the Ruth and Herman Albert Scholars Program for New Scientists in Weizmann
Institute of Science, Israel.
C.-X.L. acknowledge the support from Office of Naval Research
(Grant No. N00014-15-1-2675).
\end{acknowledgments}

%\bibliography{TopMater}


\end{document}